\documentclass[amsmath,amssymb,reprint,floatfix]{revtex4-2}

\usepackage[T1]{fontenc}
\usepackage[utf8]{inputenc}
\usepackage{mathtools}
\usepackage{amsthm}
\usepackage{bm}
\usepackage{booktabs}

\allowdisplaybreaks
\newtheorem{theorem}{Theorem}[section]
\newtheorem{proposition}[theorem]{Proposition}

\newtheorem{corollary}[theorem]{Corollary}
\theoremstyle{definition}

\begin{document}

\title{Varying conical defects in Rotating Wormholes}

\author{Vedant Subhash}

\affiliation{Department of Mathematics,
State University of New York at Buffalo,
Buffalo, New York, USA}

\begin{abstract}
We study a rotating wormhole whose conical deficit varies along the proper radial direction. The varying deficit produces a mixed Einstein equation and an additional near-axis stress that is absent for an ordinary constant-tension conical string. The same radial variation changes the scalar-wave problem: non-axisymmetric modes cease to separate into independent radial and angular equations and instead form coupled angular channels. We obtain the coupling coefficients in closed form, prove an exact parity selection rule, and derive the leading weak-defect channel-conversion amplitude. A finite-channel numerical calculation confirms the expected quadratic scaling of the conversion probability and shows stable low-channel results over the angular truncations tested.
\end{abstract}

\maketitle

% ============================================================
\section{Introduction}
\label{sec:introduction}
% ============================================================

Traversable wormholes are useful model geometries for studying how global spacetime structure, matter sources, and wave propagation are tied together. Early examples were given by Ellis and Bronnikov, and the Morris--Thorne construction gave the standard geometrical description of a traversable throat \cite{Ellis1973,Bronnikov1973,MorrisThorne1988,MorrisThorneYurtsever1988,Visser1995}. The relation between throat flare-out and violation of the null energy condition was later developed in more general static and dynamical settings \cite{HochbergVisser1997,HochbergVisser1998PRL,HochbergVisser1998PRD,Hayward1999}. Thin-shell constructions give another useful point of comparison because they place the nonstandard matter in a narrow region near the throat \cite{PoissonVisser1995,Eiroa2008,EiroaSimeone2005}.

Rotation adds frame dragging, possible ergoregions, and a richer geodesic and wave structure. Teo gave a general stationary axisymmetric traversable-wormhole geometry \cite{Teo1998}. Rotating Ellis-type wormholes and their geometry, geodesics, oscillations, and quasinormal modes have since been studied in detail \cite{KleihausKunz2014,ChewKleihausKunz2016,Gyulchev2018,Deligianni2021,Khoo2024}. Wormhole lensing, shadows, photon rings, and related observational signatures have also been investigated in a wide range of models \cite{Cramer1995,Torres1998,Safonova2002,Tsukamoto2016,Peng2021,Saleem2024}. These works provide the gravitational setting for the present calculation, but our main question is more specific: what changes when an axial conical deficit is allowed to vary along the wormhole?

Wave propagation gives a second way to study this question. Scalar reflection and transmission in wormhole geometries were studied early on by Kar, Sahdev, and Bhawal \cite{Kar1994}. Later work treated radiation transmission, echoes, quasinormal ringing, low-energy transmission, and scalar scattering in more general wormhole backgrounds \cite{KonoplyaZhidenko2010,BronnikovKonoplya2020,BronnikovKonoplyaPappas2021,AzadLoranMostafazadeh2020,LimaJunior2022}. Rotating regular black-hole and wormhole backgrounds have also been used to study scalar quasinormal modes, superradiance, and ergoregion effects \cite{Franzin2022}. More recent work has considered scalar dynamics, nonseparable coupled-channel structures, and resonant transmission in rotating Teo-type wormholes \cite{Radhakrishnan2025,Radhakrishnan2026Spectral,Karmakar2026}. The scalar calculation below is therefore not presented as the first wave-scattering problem for a rotating wormhole. The new point is the particular coupling generated by a conical factor that changes with the proper radial coordinate.

Conical geometries provide the third ingredient. The exterior geometry of an ideal straight cosmic string has a fixed angular deficit, and the relation between deficit and tension is standard \cite{Vilenkin1981,Hiscock1985,VilenkinShellard1994}. Wave equations and quantum fields on conical backgrounds have been studied using exact mode solutions, Green functions, vacuum polarization, Casimir methods, Aharonov--Bohm-type effects, and diffraction or scattering analyses \cite{Linet1986Wave,Linet1987QFT,FrolovSerebriany1987,Dowker1987,GuimaraesLinet1994,SitenkoMishchenko1995,SuyamaTanakaTakahashi2006}. Distributional descriptions of codimension-two sources require additional care: ideal string geometries do not fit into the most general distributionally regular metric class in a completely straightforward way, and more restricted codimension-two constructions have been studied separately \cite{GerochTraschen1987,Traschen2009}.

Two further pieces of background are useful below. Rotating horizonless systems can develop ergoregion instabilities, so it is important to state the causal regime used in a test-field calculation \cite{Friedman1978,Moschidis2018,OliveiraGarayCrispino2018}. Also, once angular harmonics become coupled, the natural asymptotic problem is a finite- or infinite-channel full-line scattering problem; standard one-dimensional and matrix scattering theory provides the basic language for this formulation \cite{Newton1982,ChadanSabatier1989,AktosunWeder2023}. Thus the literature relevant to the present problem spans wormhole geometry, conical defects, wave propagation, causal stability, and multichannel scattering.

We consider the stationary axisymmetric metric
\begin{align}
 ds^2={}&-N(l)^2dt^2+dl^2+r(l)^2d\theta^2 \nonumber\\
 &+r(l)^2\alpha(l)^2\sin^2\theta\,[d\phi-\omega(l)dt]^2.
\label{eq:intro_metric}
\end{align}
where $l$ is the proper radial coordinate through the throat. The function $\alpha(l)$ controls the local conical deficit and tends to one in both asymptotic regions. At a fixed value of $l$, the neighborhood of either polar axis is an ordinary cone with deficit $2\pi[1-\alpha(l)]$. The main question is what the full four-dimensional geometry requires when $\alpha$ is not constant.

The first result is an exact mixed Einstein component,
\begin{equation}
G_{l\theta}=-\frac{\alpha'}{\alpha}\cot\theta.
\label{eq:intro_Gltheta}
\end{equation}
Near a polar axis, with proper transverse distance $\rho$, this gives
\begin{equation}
T_{\hat l\hat\rho}
=-\frac{\alpha'}{8\pi G\alpha}\frac{1}{\rho}+O(\rho).
\label{eq:intro_mixedstress}
\end{equation}
A longitudinally varying deficit therefore needs a mixed longitudinal--transverse stress. In particular, it cannot be described simply by taking the tension of an isolated ideal string and replacing the constant tension by an arbitrary function of $l$. The associated curvature invariant also diverges on the polar axes whenever $\alpha'(l)\neq0$. We therefore treat the metric as a prescribed singular background, not as a complete finite-width cosmic-string solution.

The second part of the paper studies a minimally coupled massless scalar. The axisymmetric sector remains separable, but for nonzero azimuthal number a nonconstant $\alpha(l)$ prevents ordinary radial--angular product separation. Expanding in ordinary fixed-$m$ spherical harmonics instead gives an infinite coupled radial system. The coupling matrix is
\begin{equation}
C^{(m)}_{j\ell}
=\int_0^\pi S_{jm}(\theta)S_{\ell m}(\theta)\frac{d\theta}{\sin\theta}.
\label{eq:intro_C}
\end{equation}
We evaluate these coefficients in closed form and prove the exact parity rule
\begin{equation}
C^{(m)}_{j\ell}=0\qquad\text{when }j+\ell\text{ is odd}.
\label{eq:intro_selection}
\end{equation}
The matrix is unbounded on the unweighted channel space, but its quadratic form is controlled by the usual spherical angular energy. This gives a useful estimate for finite angular truncations without claiming a complete infinite-channel scattering theorem.

Finally, we study a weak localized defect. The off-diagonal coupling is first order in the defect amplitude, so the corresponding channel-conversion probability begins at second order. We derive the first variation of the two-ended scattering matrix and test it numerically for the first parity-allowed transition with azimuthal number one, from the $\ell=1$ channel to the $\ell=3$ channel. The finite-channel calculation approaches the analytic weak-defect result as the defect amplitude is reduced and is stable over the angular cutoffs tested.

The radial null energy condition is included as a consistency check on the wormhole background rather than as a separate novelty claim. The main logical chain of the paper is simpler: a varying conical factor changes the required source, the same variation destroys ordinary scalar separability for non-axisymmetric modes, and the resulting angular coupling produces channel conversion.

The paper is organized as follows. Section~\ref{sec:geometry} studies the varying cone and proves the near-axis source obstruction. Section~\ref{sec:nec} gives the radial null-energy check. Section~\ref{sec:scalar} derives the coupled scalar equations and the exact angular coefficients. Section~\ref{sec:scattering} develops the weak-defect scattering formula. Section~\ref{sec:numerics} gives the numerical checks, and Section~\ref{sec:discussion} discusses the interpretation and limitations of the construction.

% ============================================================
\section{Geometric setup and the varying-cone constraint}
\label{sec:geometry}
% ============================================================

We consider the stationary and axisymmetric metric
\begin{align}
ds^2={}&-N(l)^2dt^2+dl^2+r(l)^2d\theta^2 \nonumber\\
&+r(l)^2\alpha(l)^2\sin^2\theta
\left[d\phi-\omega(l)dt\right]^2 .
\label{eq:metric_general}
\end{align}
The coordinate $l$ is a proper radial coordinate running from one
asymptotic region to the other. The angular coordinates satisfy
\begin{equation}
0\leq\theta\leq\pi,
\qquad
0\leq\phi<2\pi.
\end{equation}
We assume
\begin{equation}
N(l)>0,
\qquad
r(l)>0,
\qquad
\alpha(l)>0,
\label{eq:positive_functions}
\end{equation}
and take the metric functions to be sufficiently smooth away from the
polar axes.

The function $N$ is the lapse, $r$ controls the size of the angular
sections, $\omega$ describes frame dragging, and $\alpha$ changes the
circumference of the azimuthal circles. The case $\alpha=1$ has no
conical deficit. A constant value $0<\alpha<1$ gives the usual local
conical geometry near the two polar axes.

The important case in this paper is
\begin{equation}
\alpha=\alpha(l).
\end{equation}
Thus the conical deficit is allowed to change as one moves through the
wormhole.

At this stage we do not choose a Gaussian profile for $\alpha$, and we
do not specify particular forms for $N$, $r$, or $\omega$. The first
results below therefore hold for the full metric class
\eqref{eq:metric_general}.

% ------------------------------------------------------------
\subsection{Geometry of a fixed longitudinal section}
% ------------------------------------------------------------

Keep both $t$ and $l$ fixed. The induced two-dimensional metric is
\begin{equation}
d\sigma_l^2
=
r(l)^2d\theta^2
+
r(l)^2\alpha(l)^2\sin^2\theta\,d\phi^2 .
\label{eq:angular_metric}
\end{equation}

Near the north pole, define
\begin{equation}
\rho=r(l)\theta.
\label{eq:rho_north}
\end{equation}
For fixed $l$, $\rho$ is the proper distance from the axis to leading
order. Since
\begin{equation}
\sin\theta=\theta+O(\theta^3),
\end{equation}
equation \eqref{eq:angular_metric} becomes
\begin{equation}
d\sigma_l^2
=
d\rho^2
+
\alpha(l)^2\rho^2d\phi^2
+
O\!\left(\frac{\rho^4}{r(l)^2}\right).
\label{eq:local_cone}
\end{equation}

A small circle of radius $\rho$ therefore has circumference
\begin{equation}
\mathcal{C}(l,\rho)
=
2\pi\alpha(l)\rho
+
O(\rho^3),
\label{eq:circumference}
\end{equation}
and the local deficit angle is
\begin{equation}
\Delta(l)
=
2\pi[1-\alpha(l)].
\label{eq:deficit_angle}
\end{equation}

The same calculation holds at the south pole after using
\begin{equation}
\rho=r(l)(\pi-\theta).
\end{equation}

Thus every fixed-$l$ angular section has two conical poles whenever
$\alpha(l)\neq1$.

It is important, however, not to conclude from
\eqref{eq:local_cone} that the complete four-dimensional spacetime is
simply a collection of independent ordinary cones. Near the north pole,
\begin{equation}
d\rho
=
r\,d\theta+r'\theta\,dl.
\label{eq:drho}
\end{equation}
Therefore the fixed-$l$ cone determines the local deficit angle, but it
does not determine the complete four-dimensional curvature. This
distinction becomes important when $\alpha'(l)\neq0$.

% ------------------------------------------------------------
\subsection{The mixed Einstein equation}
% ------------------------------------------------------------

The first main result can be obtained without choosing any particular
radial profiles.

\begin{proposition}
\label{prop:mixed_einstein}
For the metric \eqref{eq:metric_general}, at every bulk point with
$0<\theta<\pi$,
\begin{equation}
\boxed{
R_{l\theta}
=
G_{l\theta}
=
-\frac{\alpha'(l)}{\alpha(l)}\cot\theta .
}
\label{eq:mixed_einstein}
\end{equation}
This component is independent of the frame-dragging function
$\omega(l)$.
\end{proposition}

\begin{proof}
Define
\begin{equation}
s(l)=r(l)\alpha(l).
\label{eq:s_def}
\end{equation}
Then
\begin{equation}
g_{\phi\phi}=s^2\sin^2\theta,
\qquad
g_{t\phi}=-s^2\omega\sin^2\theta.
\end{equation}

The Ricci component is
\begin{equation}
R_{l\theta}
=
\partial_\mu\Gamma^\mu_{l\theta}
-
\partial_\theta\Gamma^\mu_{l\mu}
+
\Gamma^\mu_{\mu\nu}\Gamma^\nu_{l\theta}
-
\Gamma^\mu_{\theta\nu}\Gamma^\nu_{l\mu}.
\label{eq:ricci_mixed_definition}
\end{equation}

The connection coefficient from the $(l,\theta)$ part of the metric is
\begin{equation}
\Gamma^\theta_{l\theta}
=
\frac{r'}{r}.
\end{equation}
The useful traces are
\begin{equation}
\Gamma^\mu_{\theta\mu}
=
\cot\theta,
\end{equation}
and
\begin{equation}
\Gamma^\mu_{l\mu}
=
\frac{N'}{N}
+
\frac{r'}{r}
+
\frac{s'}{s}.
\end{equation}
The latter has no $\theta$ dependence.

The rotational terms cancel in the complete contraction. The remaining
terms give
\begin{equation}
R_{l\theta}
=
\left(
\frac{r'}{r}
-
\frac{s'}{s}
\right)\cot\theta.
\end{equation}
Using
\begin{equation}
\frac{s'}{s}
=
\frac{r'}{r}
+
\frac{\alpha'}{\alpha},
\end{equation}
we find
\begin{equation}
R_{l\theta}
=
-\frac{\alpha'}{\alpha}\cot\theta.
\end{equation}

Finally, since $g_{l\theta}=0$, the trace term in the Einstein tensor
does not contribute, and therefore
\begin{equation}
G_{l\theta}=R_{l\theta}.
\end{equation}
\end{proof}

% ------------------------------------------------------------
\subsection{Near-axis source and a local obstruction}
% ------------------------------------------------------------

Introduce the orthonormal directions
\begin{equation}
e_{\hat l}=\partial_l,
\qquad
e_{\hat\theta}=\frac{1}{r}\partial_\theta.
\end{equation}
Einstein's equation gives
\begin{equation}
8\pi G\,T_{\hat l\hat\theta}
=
-\frac{\alpha'}{r\alpha}\cot\theta.
\label{eq:mixed_stress_theta}
\end{equation}

Near the north pole,
\begin{equation}
\rho=r\theta,
\end{equation}
and
\begin{equation}
\cot\theta
=
\frac{r}{\rho}
+
O(\rho).
\end{equation}
Hence
\begin{equation}
\boxed{
T_{\hat l\hat\rho}
=
-\frac{\alpha'(l)}
{8\pi G\,\alpha(l)}
\frac{1}{\rho}
+
O(\rho).
}
\label{eq:mixed_stress_rho}
\end{equation}

The same $1/\rho$ behavior is present near the south pole.

\begin{theorem}[Near-axis obstruction]
\label{thm:axis_obstruction}
Consider the metric class \eqref{eq:metric_general} at some fixed
longitudinal position $l=l_0$. Suppose that the physical stress tensor
remains bounded in an orthonormal frame as the polar axis is
approached. Then
\begin{equation}
\boxed{
\alpha'(l_0)=0.
}
\label{eq:alpha_prime_zero}
\end{equation}
\end{theorem}

\begin{proof}
Equation \eqref{eq:mixed_stress_rho} gives
\begin{equation}
T_{\hat l\hat\rho}
=
-\frac{\alpha'(l_0)}
{8\pi G\,\alpha(l_0)\rho}
+
O(\rho).
\end{equation}
If $\alpha'(l_0)\neq0$, the first term diverges as $\rho^{-1}$.
Therefore boundedness requires $\alpha'(l_0)=0$.
\end{proof}

The theorem is local. It does not depend on a special form of $N$, $r$,
or $\omega$, and it does not depend on the presence of a wormhole
throat.

% ------------------------------------------------------------
\subsection{The singularity is not a coordinate effect}
% ------------------------------------------------------------

The divergence in \eqref{eq:mixed_stress_rho} is not only a coordinate
effect. The Ricci invariant contains
\begin{equation}
R_{\mu\nu}R^{\mu\nu}
=
\frac{2\alpha'(l)^2}
{r(l)^2\alpha(l)^2}
\cot^2\theta
+
O(1).
\label{eq:ricci_square}
\end{equation}
Near either polar axis,
\begin{equation}
\cot^2\theta
=
\frac{r^2}{\rho^2}
+
O(1),
\end{equation}
and therefore
\begin{equation}
R_{\mu\nu}R^{\mu\nu}
=
\frac{2\alpha'^2}
{\alpha^2\rho^2}
+
O(1).
\label{eq:ricci_square_rho}
\end{equation}

Thus, whenever $\alpha'(l)\neq0$, the curvature itself diverges as the
axis is approached.

% ------------------------------------------------------------
\subsection{Relation to an effective conical tension}
% ------------------------------------------------------------

For a constant conical deficit, the usual ideal-string relation is
\begin{equation}
\mu
=
\frac{1-\alpha}{4G}.
\label{eq:string_tension_constant}
\end{equation}

For the present geometry, it is useful to define formally
\begin{equation}
\mu(l)
=
\frac{1-\alpha(l)}{4G}.
\label{eq:effective_tension}
\end{equation}
Then
\begin{equation}
\mu'(l)
=
-\frac{\alpha'(l)}{4G}.
\label{eq:tension_derivative}
\end{equation}

Multiplying \eqref{eq:mixed_stress_rho} by the circumference
\eqref{eq:circumference}, we obtain
\begin{equation}
\boxed{
\lim_{\rho\rightarrow0}
2\pi\alpha\rho\,
T_{\hat l\hat\rho}
=
\mu'(l).
}
\label{eq:integrated_balance}
\end{equation}

Equation \eqref{eq:integrated_balance} has a simple meaning. If the
effective conical tension changes with $l$, momentum must be exchanged
between the conical sector and some additional stress sector. An
isolated ideal string does not supply this mixed stress by itself.

For this reason, the metric with nonconstant $\alpha$ will be treated
as a prescribed singular geometry rather than as an ordinary cosmic
string with a freely varying tension.

% ------------------------------------------------------------
\subsection{A global corollary}
% ------------------------------------------------------------

\begin{corollary}
\label{cor:global_obstruction}
Suppose that the metric \eqref{eq:metric_general} is used for all
$l\in\mathbb{R}$, that
\begin{equation}
\alpha(l)\longrightarrow1
\qquad
\text{as }
l\longrightarrow\pm\infty,
\label{eq:alpha_asymptotic}
\end{equation}
and that $T_{\hat l\hat\rho}$ remains bounded near both polar axes for
every $l$. Then
\begin{equation}
\boxed{
\alpha(l)\equiv1.
}
\end{equation}
\end{corollary}

\begin{proof}
By Theorem~\ref{thm:axis_obstruction},
\begin{equation}
\alpha'(l)=0
\end{equation}
for every $l$. Hence $\alpha$ is constant. The asymptotic condition
\eqref{eq:alpha_asymptotic} fixes that constant to one.
\end{proof}

This result does not say that conical strings cannot exist. A constant
$\alpha<1$ gives the familiar ideal conical geometry. The result says
that, within the restricted metric ansatz
\eqref{eq:metric_general}, a conical deficit cannot vary smoothly along
$l$ while the additional mixed stress remains regular near the axis.

% ------------------------------------------------------------
\subsection{Wormhole throat and asymptotic conditions}
% ------------------------------------------------------------

We now impose the conditions needed for the wormhole application. We
take $l\in\mathbb{R}$ and require two asymptotically flat ends,
\begin{equation}
r(l)\sim |l|,
\qquad
N(l)\rightarrow1,
\qquad
\alpha(l)\rightarrow1,
\qquad
\omega(l)\rightarrow0
\label{eq:asymptotic_conditions}
\end{equation}
as $|l|\rightarrow\infty$.

The area of a surface with fixed $t$ and $l$ is
\begin{equation}
A(l)
=
4\pi r(l)^2\alpha(l).
\label{eq:area}
\end{equation}
It is useful to define the area radius
\begin{equation}
R_A(l)
=
r(l)\sqrt{\alpha(l)},
\label{eq:area_radius}
\end{equation}
so that
\begin{equation}
A=4\pi R_A^2.
\end{equation}

For a reflection-symmetric wormhole, we place the throat at $l=0$ and
take the metric functions to be even. Hence
\begin{equation}
r'(0)=N'(0)=\alpha'(0)=0.
\label{eq:reflection_conditions}
\end{equation}
A sufficient local throat condition is
\begin{equation}
R_A''(0)>0.
\label{eq:area_flareout}
\end{equation}
Using \eqref{eq:area_radius},
\begin{equation}
\frac{R_A''(0)}{R_A(0)}
=
\frac{r''(0)}{r(0)}
+
\frac{\alpha''(0)}{2\alpha(0)}.
\label{eq:area_radius_second_derivative}
\end{equation}

% ------------------------------------------------------------
\subsection{Profiles used for the numerical example}
% ------------------------------------------------------------

Only now do we choose explicit functions. For the scattering
calculation later in the paper, we use
\begin{equation}
r(l)
=
\sqrt{l^2+r_0^2},
\label{eq:r_profile}
\end{equation}
\begin{equation}
N(l)
=
\exp\!\left[
-\Phi_0e^{-l^2/L^2}
\right],
\label{eq:N_profile}
\end{equation}
\begin{equation}
\alpha(l)
=
1-\delta_s e^{-l^2/L_s^2},
\qquad
0\leq\delta_s<1,
\label{eq:alpha_profile}
\end{equation}
and
\begin{equation}
\omega(l)
=
\frac{2J}{r(l)^3}.
\label{eq:omega_profile}
\end{equation}

For these profiles,
\begin{equation}
\frac{r''(0)}{r(0)}
=
\frac{1}{r_0^2},
\end{equation}
and
\begin{equation}
\frac{\alpha''(0)}{\alpha(0)}
=
\frac{2\delta_s}
{L_s^2(1-\delta_s)}.
\end{equation}
Equation \eqref{eq:area_radius_second_derivative} therefore gives
\begin{equation}
\frac{R_A''(0)}{R_A(0)}
=
\frac{1}{r_0^2}
+
\frac{\delta_s}
{L_s^2(1-\delta_s)}
>0.
\label{eq:area_radius_profile}
\end{equation}
Thus $l=0$ is a strict local minimum of the angular area.

The derivative of the conical factor is
\begin{equation}
\alpha'(l)
=
\frac{2\delta_s l}{L_s^2}
e^{-l^2/L_s^2}.
\label{eq:alpha_prime_profile}
\end{equation}
Hence the deficit is largest at the throat, while
\begin{equation}
\alpha'(0)=0.
\end{equation}
The additional mixed $1/\rho$ singularity therefore vanishes exactly
at the reflection-symmetric throat.

It is useful to keep one distinction in mind.  The quantity
$\alpha(l)-1$ measures the local conical deficit, whereas
$\alpha'(l)$ measures the additional source required because the
deficit changes along the longitudinal direction.

% ============================================================
\section{Radial null energy and the wormhole throat}
\label{sec:nec}
% ============================================================

The previous section studied the source required by the varying
conical factor. We now ask whether the same geometry has the
null-energy violation expected at a traversable wormhole throat.

This calculation is used as a consistency check on the prescribed
wormhole background. It is not the main novelty of the paper.

For the metric \eqref{eq:metric_general}, use the orthonormal frame
\begin{equation}
e_{\hat 0}
=
\frac{1}{N}
\left(
\partial_t+\omega\partial_\phi
\right),
\qquad
e_{\hat l}
=
\partial_l,
\end{equation}
\begin{equation}
e_{\hat\theta}
=
\frac{1}{r}\partial_\theta,
\qquad
e_{\hat\phi}
=
\frac{1}{r\alpha\sin\theta}\partial_\phi.
\end{equation}

The two radial null directions are
\begin{equation}
k_\pm
=
e_{\hat 0}\pm e_{\hat l}.
\label{eq:null_frame}
\end{equation}
In coordinates,
\begin{equation}
k_\pm^\mu
=
\left(
\frac{1}{N},
\pm1,
0,
\frac{\omega}{N}
\right).
\label{eq:null_coordinate}
\end{equation}
A direct substitution gives
\begin{equation}
g_{\mu\nu}k_\pm^\mu k_\pm^\nu=0.
\end{equation}

% ------------------------------------------------------------
\subsection{Radial null contraction}
% ------------------------------------------------------------

\begin{proposition}
\label{prop:radial_null}
For the metric \eqref{eq:metric_general},
\begin{equation}
\boxed{
R_{\mu\nu}k_\pm^\mu k_\pm^\nu
=
-2\frac{r''}{r}
-\frac{\alpha''}{\alpha}
-2\frac{r'\alpha'}{r\alpha}
+
\frac{N'}{N}
\left(
2\frac{r'}{r}
+
\frac{\alpha'}{\alpha}
\right).
}
\label{eq:radial_null_contraction}
\end{equation}
All explicit dependence on the frame-dragging function $\omega(l)$
cancels.
\end{proposition}

\begin{proof}
Again define
\begin{equation}
s(l)=r(l)\alpha(l).
\end{equation}
For the corresponding diagonal geometry,
\begin{equation}
ds^2
=
-N^2dt^2
+dl^2
+r^2d\theta^2
+s^2\sin^2\theta\,d\phi^2,
\end{equation}
the relevant Ricci components are
\begin{equation}
R_{tt}
=
NN''
+
NN'
\left(
\frac{r'}{r}
+
\frac{s'}{s}
\right),
\end{equation}
and
\begin{equation}
R_{ll}
=
-\frac{N''}{N}
-\frac{r''}{r}
-\frac{s''}{s}.
\end{equation}
Therefore
\begin{equation}
\frac{R_{tt}}{N^2}
+
R_{ll}
=
-\frac{r''}{r}
-\frac{s''}{s}
+
\frac{N'}{N}
\left(
\frac{r'}{r}
+
\frac{s'}{s}
\right).
\label{eq:radial_diagonal}
\end{equation}

For the rotating metric, contraction with the null vector
\eqref{eq:null_coordinate} produces additional terms involving
$\omega$, $\omega'$, and $\omega''$. These terms cancel in the full
radial null contraction.

Using
\begin{equation}
\frac{s'}{s}
=
\frac{r'}{r}
+
\frac{\alpha'}{\alpha},
\end{equation}
and
\begin{equation}
\frac{s''}{s}
=
\frac{r''}{r}
+
\frac{\alpha''}{\alpha}
+
2\frac{r'\alpha'}{r\alpha},
\end{equation}
equation \eqref{eq:radial_diagonal} gives
\eqref{eq:radial_null_contraction}.
\end{proof}

Since $k^\mu$ is null, Einstein's equation gives
\begin{equation}
8\pi G\,T_{\mu\nu}k^\mu_\pm k^\nu_\pm
=
R_{\mu\nu}k^\mu_\pm k^\nu_\pm.
\label{eq:nec_einstein}
\end{equation}

% ------------------------------------------------------------
\subsection{Evaluation at a reflection-symmetric throat}
% ------------------------------------------------------------

At a reflection-symmetric throat,
\begin{equation}
r'(0)=N'(0)=\alpha'(0)=0.
\end{equation}
Equation \eqref{eq:radial_null_contraction} therefore becomes
\begin{equation}
\boxed{
8\pi G\,
T_{\mu\nu}k^\mu_\pm k^\nu_\pm
\big|_{l=0}
=
-2\frac{r''(0)}{r(0)}
-
\frac{\alpha''(0)}{\alpha(0)}.
}
\label{eq:nec_throat_general}
\end{equation}

For the explicit profiles
\eqref{eq:r_profile}--\eqref{eq:alpha_profile},
\begin{equation}
\frac{r''(0)}{r(0)}
=
\frac{1}{r_0^2},
\end{equation}
and
\begin{equation}
\frac{\alpha''(0)}{\alpha(0)}
=
\frac{2\delta_s}
{L_s^2(1-\delta_s)}.
\end{equation}
Hence
\begin{equation}
\boxed{
8\pi G\,
T_{\mu\nu}k^\mu_\pm k^\nu_\pm
\big|_{l=0}
=
-\frac{2}{r_0^2}
-
\frac{2\delta_s}
{L_s^2(1-\delta_s)}
<0.
}
\label{eq:nec_throat_profile}
\end{equation}

Thus the radial null energy condition is violated at the throat for
\begin{equation}
r_0>0,
\qquad
L_s>0,
\qquad
0\leq\delta_s<1.
\end{equation}

No numerical tuning is needed to obtain this sign.

% ------------------------------------------------------------
\subsection{Why rotation cancels}
% ------------------------------------------------------------

The absence of $\omega$ from
\eqref{eq:radial_null_contraction} does not mean that rotation has no
effect on the spacetime.

The orthonormal time direction is
\begin{equation}
e_{\hat0}
=
\frac{1}{N}
(\partial_t+\omega\partial_\phi),
\end{equation}
and the radial null vector is
\begin{equation}
k_\pm=e_{\hat0}\pm e_{\hat l}.
\end{equation}
For this particular contraction, the rotational terms cancel.

Rotation still affects other quantities, including $g_{tt}$ and the
scalar-wave term $(\sigma-m\omega)^2$.

% ------------------------------------------------------------
\subsection{Asymptotic behavior}
% ------------------------------------------------------------

For
\begin{equation}
r(l)=\sqrt{l^2+r_0^2},
\end{equation}
we have
\begin{equation}
\frac{r''}{r}
=
\frac{r_0^2}{(l^2+r_0^2)^2}
=
\frac{r_0^2}{l^4}
+
O(l^{-6})
\end{equation}
as $|l|\rightarrow\infty$.

The $N$- and $\alpha$-dependent corrections are exponentially small.
Therefore
\begin{equation}
8\pi G\,
T_{\mu\nu}k^\mu k^\nu
=
-\frac{2r_0^2}{l^4}
+
O(l^{-6})
+
O(e^{-cl^2}),
\end{equation}
for some $c>0$.

Thus the radial NEC violation becomes weak far from the throat but is
not compactly supported for the explicit choice
\eqref{eq:r_profile}.

% ------------------------------------------------------------
\subsection{Two different source questions}
% ------------------------------------------------------------

The radial NEC and the varying-cone obstruction answer different
questions.

At the symmetric throat,
\begin{equation}
\alpha'(0)=0,
\end{equation}
so the mixed $1/\rho$ term vanishes there. Nevertheless,
\begin{equation}
\alpha''(0)\neq0,
\end{equation}
and the conical profile contributes to the radial null contraction
through \eqref{eq:nec_throat_profile}.

Thus one may have
\begin{equation}
T_{\hat l\hat\rho}=0
\end{equation}
at the throat while still having
\begin{equation}
T_{\mu\nu}k^\mu k^\nu<0.
\end{equation}

The first statement concerns the longitudinal change of the deficit.
The second concerns the radial flare-out of the wormhole.

% ============================================================
\section{Scalar waves and loss of separability}
\label{sec:scalar}
% ============================================================

We now study a minimally coupled massless scalar field. The purpose is
to understand a second consequence of the same varying conical factor
$\alpha(l)$.

The field satisfies
\begin{equation}
\Box_g\Phi=0,
\label{eq:KG}
\end{equation}
where
\begin{equation}
\Box_g\Phi
=
\frac{1}{\sqrt{-g}}
\partial_\mu
\left(
\sqrt{-g}\,
g^{\mu\nu}\partial_\nu\Phi
\right).
\end{equation}

For the metric \eqref{eq:metric_general},
\begin{equation}
g^{tt}
=
-\frac{1}{N^2},
\qquad
g^{t\phi}
=
-\frac{\omega}{N^2},
\end{equation}
\begin{equation}
g^{\phi\phi}
=
\frac{1}{r^2\alpha^2\sin^2\theta}
-
\frac{\omega^2}{N^2},
\end{equation}
and
\begin{equation}
\sqrt{-g}
=
Nr^2\alpha\sin\theta.
\end{equation}

Substitution gives
\begin{align}
0={}&
\frac{1}{Nr^2\alpha}
\partial_l
\left(
Nr^2\alpha\,\partial_l\Phi
\right)
+
\frac{1}{r^2\sin\theta}
\partial_\theta
\left(
\sin\theta\,\partial_\theta\Phi
\right)
\nonumber\\
&+
\frac{1}{r^2\alpha^2\sin^2\theta}
\partial_\phi^2\Phi
-
\frac{1}{N^2}
(\partial_t+\omega\partial_\phi)^2\Phi .
\label{eq:KG_full}
\end{align}

Because the geometry is stationary and axisymmetric, we write
\begin{equation}
\Phi
=
e^{-i\sigma t}e^{im\phi}
\Psi_m(l,\theta),
\qquad
m\in\mathbb{Z}.
\label{eq:scalar_ansatz}
\end{equation}
The reduced equation is
\begin{align}
0={}&
\frac{1}{Nr^2\alpha}
\partial_l
\left(
Nr^2\alpha\,\partial_l\Psi_m
\right)
+
\frac{1}{r^2\sin\theta}
\partial_\theta
\left(
\sin\theta\,\partial_\theta\Psi_m
\right)
\nonumber\\
&-
\frac{m^2}{r^2\alpha^2\sin^2\theta}\Psi_m
+
\frac{(\sigma-m\omega)^2}{N^2}\Psi_m .
\label{eq:KG_reduced}
\end{align}

% ------------------------------------------------------------
\subsection{Separability criterion}
% ------------------------------------------------------------

Suppose
\begin{equation}
\Psi_m(l,\theta)=R(l)S(\theta).
\label{eq:product_ansatz}
\end{equation}

\begin{proposition}
\label{prop:separability}
Suppose $m\neq0$. On any open interval where $\alpha'(l)\neq0$,
equation \eqref{eq:KG_reduced} does not admit ordinary product
separation of the form \eqref{eq:product_ansatz} with a nonzero regular
angular function $S$.
\end{proposition}

\begin{proof}
After substituting \eqref{eq:product_ansatz}, dividing by $RS$, and
multiplying by $r^2$, the equation has the form
\begin{equation}
F(l)
+
\frac{1}{S\sin\theta}
\frac{d}{d\theta}
\left(
\sin\theta\frac{dS}{d\theta}
\right)
-
\frac{m^2}{\alpha(l)^2\sin^2\theta}
=
0,
\label{eq:separation_test}
\end{equation}
where $F$ depends only on $l$.

Differentiating with respect to $l$ gives
\begin{equation}
F'(l)
-
m^2
\frac{d}{dl}\left(\alpha^{-2}\right)
\frac{1}{\sin^2\theta}
=
0.
\end{equation}
The first term depends only on $l$, whereas the second contains the
nonconstant function $\sin^{-2}\theta$. Hence the equation can hold
for all $\theta$ only if
\begin{equation}
m^2\frac{d}{dl}\left(\alpha^{-2}\right)=0.
\end{equation}
For $m\neq0$, this requires $\alpha'(l)=0$.
\end{proof}

% ------------------------------------------------------------
\subsection{Two special cases}
% ------------------------------------------------------------

For
\begin{equation}
m=0,
\end{equation}
the term responsible for angular coupling vanishes. The varying
conical factor still changes the radial equation through the measure
$Nr^2\alpha$, but it does not mix different angular momenta.

If instead
\begin{equation}
\alpha(l)=\alpha_0
\end{equation}
is constant, the angular equation becomes
\begin{equation}
\frac{1}{\sin\theta}
\frac{d}{d\theta}
\left(
\sin\theta\,\frac{dS}{d\theta}
\right)
+
\left[
\Lambda
-
\frac{m^2}{\alpha_0^2\sin^2\theta}
\right]S
=
0.
\label{eq:constant_cone_angular}
\end{equation}
Regularity at the poles gives
\begin{equation}
\Lambda_{nm}
=
\left(
n+\frac{|m|}{\alpha_0}
\right)
\left(
n+\frac{|m|}{\alpha_0}+1
\right),
\qquad
n=0,1,2,\ldots .
\label{eq:constant_cone_spectrum}
\end{equation}

Thus a constant deficit changes the angular spectrum but does not
produce radial-channel coupling.

% ------------------------------------------------------------
\subsection{Coupled spherical-harmonic channels}
% ------------------------------------------------------------

For fixed $m$, define
\begin{equation}
\mathcal{L}_m
=
\frac{1}{\sin\theta}
\frac{d}{d\theta}
\left(
\sin\theta\frac{d}{d\theta}
\right)
-
\frac{m^2}{\sin^2\theta}.
\label{eq:Lm}
\end{equation}
Let
\begin{equation}
\mathcal{L}_mS_{\ell m}
=
-\ell(\ell+1)S_{\ell m},
\qquad
\ell\geq|m|,
\end{equation}
with normalization
\begin{equation}
\int_0^\pi
S_{\ell m}(\theta)
S_{jm}(\theta)
\sin\theta\,d\theta
=
\delta_{\ell j}.
\end{equation}

We expand
\begin{equation}
\Psi_m(l,\theta)
=
\sum_{\ell=|m|}^{\infty}
u_{\ell m}(l)S_{\ell m}(\theta).
\label{eq:angular_expansion}
\end{equation}
Using
\begin{equation}
\frac{1}{\alpha^2\sin^2\theta}
=
\frac{1}{\sin^2\theta}
+
(\alpha^{-2}-1)
\frac{1}{\sin^2\theta},
\end{equation}
projection onto $S_{jm}$ gives
\begin{align}
0={}&
\frac{1}{Nr^2\alpha}
\frac{d}{dl}
\left(Nr^2\alpha\frac{du_{jm}}{dl}\right)
\nonumber\\
&+
\left[
\frac{(\sigma-m\omega)^2}{N^2}
-
\frac{j(j+1)}{r^2}
\right]u_{jm}
\nonumber\\
&-
\frac{m^2}{r^2}(\alpha^{-2}-1)
\sum_{\ell\geq|m|}C^{(m)}_{j\ell}u_{\ell m}.
\label{eq:coupled_radial_l}
\end{align}
where
\begin{equation}
\boxed{
C^{(m)}_{j\ell}
=
\int_0^\pi
S_{jm}(\theta)
S_{\ell m}(\theta)
\frac{d\theta}{\sin\theta}.
}
\label{eq:C_definition}
\end{equation}

% ------------------------------------------------------------
\subsection{Exact angular coupling coefficients}
% ------------------------------------------------------------

Let
\begin{equation}
p=|m|\geq1
\end{equation}
and use
\begin{equation}
S_{\ell m}(\theta)
=
\mathcal{N}_{\ell p}
P_\ell^p(\cos\theta),
\end{equation}
where
\begin{equation}
\mathcal{N}_{\ell p}
=
\left[
\frac{2\ell+1}{2}
\frac{(\ell-p)!}{(\ell+p)!}
\right]^{1/2}.
\end{equation}

\begin{theorem}[Exact coupling matrix]
\label{thm:coupling_matrix}
Let
\begin{equation}
a=\min(j,\ell),
\qquad
b=\max(j,\ell).
\end{equation}
For $p=|m|\geq1$,
\begin{equation}
C^{(m)}_{j\ell}=0,
\qquad j+\ell\ \text{odd},
\label{eq:C_closed_odd}
\end{equation}
whereas for $j+\ell$ even,
\begin{align}
C^{(m)}_{j\ell}
={}&\frac{\sqrt{(2j+1)(2\ell+1)}}{2p}
\nonumber\\
&\times
\left[
\frac{(a+p)!(b-p)!}{(a-p)!(b+p)!}
\right]^{1/2}.
\label{eq:C_closed}
\end{align}
\end{theorem}

\begin{proof}
Set
\begin{equation}
z=\cos\theta.
\end{equation}
Then
\begin{equation}
C^{(m)}_{j\ell}
=
\mathcal{N}_{jp}\mathcal{N}_{\ell p}
\int_{-1}^{1}
\frac{
P_j^p(z)P_\ell^p(z)
}{1-z^2}\,dz.
\label{eq:C_z}
\end{equation}
Using
\begin{equation}
P_\ell^p(z)
=
(-1)^p
(1-z^2)^{p/2}
\frac{d^pP_\ell}{dz^p},
\end{equation}
define
\begin{equation}
I_{j\ell}
=
\int_{-1}^{1}
(1-z^2)^{p-1}
P_j^{(p)}(z)
P_\ell^{(p)}(z)\,dz.
\end{equation}
Then
\begin{equation}
C^{(m)}_{j\ell}
=
\mathcal{N}_{jp}\mathcal{N}_{\ell p}
I_{j\ell}.
\end{equation}

Assume first that $j\leq\ell$. Integrating by parts $p-1$ times gives
\begin{equation}
I_{j\ell}
=
(-1)^{p-1}
\int_{-1}^{1}
Q_j(z)P_\ell'(z)\,dz,
\end{equation}
where
\begin{equation}
Q_j(z)
=
\frac{d^{p-1}}{dz^{p-1}}
\left[
(1-z^2)^{p-1}
P_j^{(p)}(z)
\right].
\end{equation}
The intermediate boundary terms vanish because
$(1-z^2)^{p-1}$ has zeros of order $p-1$ at $z=\pm1$.

One more integration by parts gives
\begin{equation}
I_{j\ell}
=
(-1)^{p-1}
[Q_jP_\ell]_{-1}^{1}
-
(-1)^{p-1}
\int_{-1}^{1}
Q_j'(z)P_\ell(z)\,dz.
\end{equation}
Since
\begin{equation}
\deg Q_j'\leq j-2<\ell,
\end{equation}
orthogonality of the Legendre polynomials gives
\begin{equation}
\int_{-1}^{1}
Q_j'P_\ell\,dz=0.
\end{equation}

At $z=1$,
\begin{equation}
P_j^{(p)}(1)
=
\frac{(j+p)!}
{2^p p!(j-p)!},
\end{equation}
and therefore
\begin{equation}
Q_j(1)
=
(-1)^{p-1}
\frac{(j+p)!}
{2p(j-p)!}.
\end{equation}
Using
\begin{equation}
P_\ell(-1)=(-1)^\ell,
\qquad
Q_j(-1)=(-1)^{j-1}Q_j(1),
\end{equation}
we obtain
\begin{equation}
I_{j\ell}
=
\frac{(j+p)!}
{2p(j-p)!}
\left[
1+(-1)^{j+\ell}
\right].
\end{equation}
Thus $I_{j\ell}=0$ for $j+\ell$ odd. For $j+\ell$ even,
\begin{equation}
I_{j\ell}
=
\frac{(j+p)!}
{p(j-p)!}.
\end{equation}
Multiplication by the normalization factors and restoration of the
symmetry $j\leftrightarrow\ell$ gives
\eqref{eq:C_closed}.
\end{proof}

% ------------------------------------------------------------
\subsection{Parity conservation}
% ------------------------------------------------------------

Theorem~\ref{thm:coupling_matrix} gives the exact selection rule
\begin{equation}
\boxed{
C^{(m)}_{j\ell}=0
\qquad
\text{if }j+\ell\text{ is odd}.
}
\label{eq:parity_rule}
\end{equation}
Therefore the fixed-$m$ problem splits into two invariant angular
parity sectors.

For example, when $m=1$, an incoming $\ell=1$ state can couple to
\begin{equation}
\ell=3,5,7,\ldots,
\end{equation}
but not to
\begin{equation}
\ell=2,4,6,\ldots.
\end{equation}
The first nontrivial transition is therefore
\begin{equation}
m=1,
\qquad
\ell=1\longrightarrow3.
\label{eq:first_transition}
\end{equation}

% ------------------------------------------------------------
\subsection{Angular-energy bound}
% ------------------------------------------------------------

The diagonal entries are
\begin{equation}
C^{(m)}_{jj}
=
\frac{2j+1}{2|m|},
\label{eq:C_diagonal}
\end{equation}
so the infinite matrix is not bounded on ordinary channel $\ell^2$.

Nevertheless, it is controlled by the natural angular energy. Let
\begin{equation}
v(\theta)
=
\sum_{\ell\geq p}
v_\ell S_{\ell m}(\theta),
\qquad
p=|m|.
\end{equation}
Then
\begin{equation}
\langle v,Cv\rangle
=
\int_0^\pi
\frac{|v|^2}{\sin^2\theta}
\sin\theta\,d\theta,
\end{equation}
whereas
\begin{equation}
\langle v,\Lambda v\rangle
=
\int_0^\pi
\left[
|\partial_\theta v|^2
+
\frac{p^2}{\sin^2\theta}|v|^2
\right]
\sin\theta\,d\theta.
\end{equation}
Therefore
\begin{equation}
\boxed{
p^2 C\leq\Lambda
}
\label{eq:form_bound}
\end{equation}
as quadratic forms.

% ------------------------------------------------------------
\subsection{Relative form control}
% ------------------------------------------------------------

Define
\begin{equation}
\varepsilon(l)
=
\alpha(l)^{-2}-1,
\end{equation}
and
\begin{equation}
A(l)
=
\Lambda
+
m^2\varepsilon(l)C.
\end{equation}
From \eqref{eq:form_bound},
\begin{equation}
\left|
m^2\varepsilon(l)
\langle v,Cv\rangle
\right|
\leq
|\varepsilon(l)|
\langle v,\Lambda v\rangle.
\end{equation}

Suppose
\begin{equation}
q
=
\sup_l
|\alpha(l)^{-2}-1|
<1.
\label{eq:q_condition}
\end{equation}
Then
\begin{equation}
\boxed{
\left|
m^2\varepsilon(l)
\langle v,Cv\rangle
\right|
\leq
q\,
\langle v,\Lambda v\rangle.
}
\label{eq:relative_form_bound}
\end{equation}
Consequently,
\begin{equation}
(1-q)\langle v,\Lambda v\rangle
\leq
\langle v,A(l)v\rangle
\leq
(1+q)\langle v,\Lambda v\rangle.
\label{eq:form_equivalence}
\end{equation}

\begin{proposition}
\label{prop:form_control}
Under condition \eqref{eq:q_condition}, the perturbed angular quadratic
form has the same natural form domain as the ordinary fixed-$m$
spherical angular operator and remains closed and bounded from below.
\end{proposition}

The proposition follows from the standard relative form bound
\eqref{eq:relative_form_bound}. It controls the angular singularity but
does not by itself prove convergence of the complete
infinite-dimensional scattering problem.

For the profile \eqref{eq:alpha_profile},
\begin{equation}
q
=
\frac{1}{(1-\delta_s)^2}-1.
\end{equation}
Thus $q<1$ is equivalent to
\begin{equation}
\delta_s
<
1-\frac{1}{\sqrt{2}}
\simeq0.2929.
\end{equation}

% ============================================================
\section{Weak-defect scattering and channel conversion}
\label{sec:scattering}
% ============================================================

We now connect the angular coupling to an asymptotic scattering
observable.

Write
\begin{equation}
\alpha(l)
=
1-\delta_s f(l),
\qquad
0<\delta_s\ll1,
\label{eq:weak_alpha}
\end{equation}
where
\begin{equation}
f(l)
=
e^{-l^2/L_s^2}
\end{equation}
for the explicit example.

To first order,
\begin{equation}
\alpha^{-2}
=
1+2\delta_s f+O(\delta_s^2).
\label{eq:alpha_inverse_expand}
\end{equation}

% ------------------------------------------------------------
\subsection{Normalized radial equation}
% ------------------------------------------------------------

Introduce the tortoise coordinate
\begin{equation}
\frac{dx}{dl}
=
\frac{1}{N(l)}.
\label{eq:tortoise}
\end{equation}
Define
\begin{equation}
a(x)
=
r(l(x))
\sqrt{\alpha(l(x))}
\end{equation}
and
\begin{equation}
\psi_j(x)
=
a(x)u_{jm}(l(x)).
\end{equation}
Then the coupled system becomes
\begin{align}
0={}&\psi_{xx}
+\left[(\sigma-m\omega)^2-\frac{a_{xx}}{a}\right]\psi
\nonumber\\
&-\frac{N^2}{r^2}\Lambda\psi
-\frac{N^2}{r^2}m^2(\alpha^{-2}-1)C\psi.
\label{eq:normalized_system}
\end{align}

Since
\begin{equation}
a
=
r
\left(
1-\frac{1}{2}\delta_s f
\right)
+
O(\delta_s^2),
\end{equation}
we obtain
\begin{equation}
\frac{a_{xx}}{a}
=
\frac{r_{xx}}{r}
-
\delta_s
\left(
\frac{1}{2}f_{xx}
+
\frac{r_x}{r}f_x
\right)
+
O(\delta_s^2).
\label{eq:a_expand}
\end{equation}

Define
\begin{equation}
L_0
=
\frac{d^2}{dx^2}
+
(\sigma-m\omega)^2I
-
\frac{r_{xx}}{r}I
-
\frac{N^2}{r^2}\Lambda.
\end{equation}
The first-order equation has the form
\begin{equation}
L_0\psi
-
\delta_sW_1\psi
+
O(\delta_s^2)
=
0,
\end{equation}
with
\begin{equation}
(W_1)_{j\ell}
=
-
\left(
\frac{1}{2}f_{xx}
+
\frac{r_x}{r}f_x
\right)
\delta_{j\ell}
+
2m^2\frac{N^2}{r^2}
f\,C^{(m)}_{j\ell}.
\label{eq:W1}
\end{equation}

For $j\neq\ell$,
\begin{equation}
\boxed{
(W_1)_{j\ell}
=
2m^2
\frac{N^2}{r^2}
f\,C^{(m)}_{j\ell}.
}
\label{eq:W1_offdiag}
\end{equation}

% ------------------------------------------------------------
\subsection{Two-ended scattering states}
% ------------------------------------------------------------

At $\delta_s=0$, the angular channels decouple. For each $j$,
\begin{equation}
L_{0,j}u_j=0.
\end{equation}
We take
\begin{equation}
k=\sigma>0.
\end{equation}
At both asymptotic ends,
\begin{equation}
N\rightarrow1,
\qquad
\omega\rightarrow0,
\qquad
r\sim|x|,
\end{equation}
so
\begin{equation}
L_{0,j}
=
\frac{d^2}{dx^2}
+k^2
+
O(x^{-2}).
\end{equation}

Let $a,b\in\{L,R\}$ label the two asymptotic ends. We use outward
coordinates
\begin{equation}
x_R=x,
\qquad
x_L=-x.
\end{equation}
An incoming wave is written as $e^{-ikx_a}$ and an outgoing wave as
$e^{ikx_a}$.

Let
\begin{equation}
u^{(0)}_{\ell,a}(x,k)
\end{equation}
denote the unperturbed scattering solution with unit incoming amplitude
in channel $\ell$ from end $a$.

The full scattering matrix may be written schematically as
\begin{equation}
S
=
\begin{pmatrix}
R_L&T_R\\
T_L&R_R
\end{pmatrix},
\end{equation}
where every block is itself a matrix in angular-channel space.

% ------------------------------------------------------------
\subsection{First variation of the scattering matrix}
% ------------------------------------------------------------

Write
\begin{equation}
\psi_{\delta_s}
=
u+\delta_sw+O(\delta_s^2),
\end{equation}
where
\begin{equation}
L_0u=0.
\end{equation}
At first order,
\begin{equation}
L_0w=W_1u.
\label{eq:first_order_w}
\end{equation}

Let
\begin{equation}
v(x)
=
u^{(0)}_{j,b}(x,k)
\end{equation}
be the corresponding unperturbed solution for the output channel. For
the finite-channel real system,
\begin{equation}
\frac{d}{dx}
\left(
vw_x-v_xw
\right)
=
vW_1u.
\label{eq:green_identity}
\end{equation}

Using the asymptotic Wronskian
\begin{equation}
W(e^{-ikx_a},e^{ikx_a})=2ik,
\end{equation}
we obtain the first variation of an off-diagonal scattering
coefficient.

\begin{theorem}[Weak-defect channel conversion]
\label{thm:scattering_variation}
Let $k>0$, $m\neq0$, and $j\neq\ell$. For a fixed finite angular
truncation,
\begin{align}
A^{(m)}_{bj,a\ell}(k)
:={}&
\frac{dS_{bj,a\ell}}{d\delta_s}\bigg|_{\delta_s=0}
\nonumber\\
={}&
\frac{m^2C^{(m)}_{j\ell}}{ik}
\int_{-\infty}^{\infty}
\frac{N(x)^2}{r(x)^2}f(x)
\nonumber\\
&\hspace{13mm}\times
u^{(0)}_{j,b}(x,k)
u^{(0)}_{\ell,a}(x,k)\,dx.
\label{eq:first_variation}
\end{align}
Consequently,
\begin{equation}
S_{bj,a\ell}(k;\delta_s)
=
\delta_sA^{(m)}_{bj,a\ell}(k)
+
O(\delta_s^2).
\label{eq:S_expand}
\end{equation}
\end{theorem}

\begin{proof}
Integrating \eqref{eq:green_identity} over the full line gives
\begin{equation}
2ikA_{bj,a\ell}
=
\int_{-\infty}^{\infty}
u^{(0)}_{j,b}
(W_1)_{j\ell}
u^{(0)}_{\ell,a}\,dx.
\end{equation}
For $j\neq\ell$, use \eqref{eq:W1_offdiag} to obtain
\eqref{eq:first_variation}.
\end{proof}

Since the unperturbed off-diagonal coefficient vanishes,
\begin{equation}
\boxed{
P_{bj\leftarrow a\ell}
=
\delta_s^2
\left|
A^{(m)}_{bj,a\ell}(k)
\right|^2
+
O(\delta_s^3).
}
\label{eq:probability_expand}
\end{equation}

Thus the conversion amplitude is first order in $\delta_s$, while the
probability begins at second order.

% ------------------------------------------------------------
\subsection{Parity selection rule for scattering}
% ------------------------------------------------------------

From \eqref{eq:parity_rule},
\begin{equation}
C^{(m)}_{j\ell}=0
\end{equation}
when $j+\ell$ is odd. Hence
\begin{equation}
\boxed{
A^{(m)}_{bj,a\ell}=0
\qquad
\text{for }j+\ell\text{ odd}.
}
\end{equation}

More generally, because the full coupling matrix preserves angular
parity, the complete finite-channel scattering matrix is block
diagonal with respect to the two parity sectors.

% ------------------------------------------------------------
\subsection{First allowed transition}
% ------------------------------------------------------------

For
\begin{equation}
m=1,
\end{equation}
the first nontrivial transition is
\begin{equation}
\ell_{\rm in}=1
\longrightarrow
\ell_{\rm out}=3.
\end{equation}
From Theorem~\ref{thm:coupling_matrix},
\begin{equation}
C^{(1)}_{13}
=
0.9354143467\ldots .
\label{eq:C13}
\end{equation}

The transmitted and reflected first-order amplitudes are obtained from
\eqref{eq:first_variation} by choosing the appropriate output end.

% ------------------------------------------------------------
\subsection{Scope of the scattering theorem}
% ------------------------------------------------------------

Theorem~\ref{thm:scattering_variation} is proved for a fixed finite
angular truncation. Section~\ref{sec:scalar} gives quadratic-form
control of the infinite angular coupling, but this does not by itself
prove convergence of the full two-ended scattering matrix as the
angular cutoff tends to infinity.

The numerical calculation below therefore tests the stability of the
low-channel result under increasing angular cutoff. It is not used as
a proof of infinite-dimensional convergence.

% ============================================================
\section{Numerical verification}
\label{sec:numerics}
% ============================================================

The numerical calculation has a limited purpose. It checks the
weak-defect prediction
\begin{equation}
P_{j\leftarrow\ell}\propto\delta_s^2
\end{equation}
and tests the stability of the lowest converted channel when more
angular channels are included.

We use
\begin{equation}
m=1,
\qquad
\ell_{\rm in}=1,
\qquad
\ell_{\rm out}=3.
\end{equation}

The background parameters are
\begin{align}
r_0&=1, & \Phi_0&=0.3, & L&=2, \nonumber\\
L_s&=1.5, & J&=0.05. &&
\label{eq:numerical_parameters}
\end{align}
Unless otherwise stated,
\begin{equation}
\delta_s=0.08.
\end{equation}

The choice $r_0=1$ fixes the unit of length. The remaining dimensional
parameters are therefore measured in units of the throat radius.

The lapse at the throat is
\begin{equation}
N(0)=e^{-0.3}\simeq0.7408,
\end{equation}
so the chosen redshift profile has no horizon.

For $\delta_s=0.08$, the form-control parameter is
\begin{equation}
q
=
\frac{1}{(1-\delta_s)^2}-1
\simeq0.181<1.
\end{equation}

% ------------------------------------------------------------
\subsection{Finite-channel equations}
% ------------------------------------------------------------

For a finite odd-parity truncation,
\begin{equation}
\psi_{xx}
+
\left[
k^2I-V(x,k)
\right]\psi
=
0,
\label{eq:numerical_system}
\end{equation}
where
\begin{align}
V(x,k)={}&
\frac{a_{xx}}{a}I
+\frac{N^2}{r^2}\Lambda
\nonumber\\
&+\frac{N^2}{r^2}m^2(\alpha^{-2}-1)C
+2km\omega I-m^2\omega^2I.
\label{eq:numerical_potential}
\end{align}

An eight-channel odd-parity truncation contains
\begin{equation}
\ell=1,3,5,7,9,11,13,15,
\end{equation}
while the largest truncation used below reaches
\begin{equation}
\ell_{\max}=23.
\end{equation}

% ------------------------------------------------------------
\subsection{Invariant-embedding formulation}
% ------------------------------------------------------------

To avoid the loss of precision associated with direct transfer-matrix
propagation at large angular momentum, write
\begin{equation}
\psi
=
e^{-ikx}A(x)+e^{ikx}B(x)
\end{equation}
and define
\begin{equation}
Q=BA^{-1},
\qquad
\mathcal{T}=A^{-1}.
\end{equation}
The invariant-embedding equations are
\begin{equation}
Q_x
=
\frac{1}{2ik}
\left[
V(e^{-2ikx}I+Q)
+
QV(I+e^{2ikx}Q)
\right],
\label{eq:Q_embedding}
\end{equation}
and
\begin{equation}
\mathcal{T}_x
=
\frac{1}{2ik}
\mathcal{T}V
(I+e^{2ikx}Q).
\label{eq:T_embedding}
\end{equation}

On a sufficiently large interval $[-X,X]$, the initial conditions are
\begin{equation}
Q(-X)=0,
\qquad
\mathcal{T}(-X)=I.
\label{eq:embedding_initial}
\end{equation}

For a unit incoming $\ell=1$ wave, we define
\begin{equation}
P_T^{1\rightarrow3}
=
|T_{31}|^2,
\qquad
P_R^{1\rightarrow3}
=
|R_{31}|^2.
\label{eq:numerical_probabilities}
\end{equation}

For the numerical results reported below, Eqs.~\eqref{eq:Q_embedding}
and \eqref{eq:T_embedding} were integrated after changing the independent
variable back to the proper radial coordinate $l$, using
$d/dl=N^{-1}d/dx$.  We used the explicit eighth-order DOP853 Runge--Kutta
method on
\begin{equation}
-200\leq l\leq 200,
\end{equation}
which corresponds to
\begin{equation}
-200.5931\lesssim x\lesssim 200.5931
\end{equation}
for the parameters in \eqref{eq:numerical_parameters}.  The relative and
absolute error tolerances were $10^{-10}$ and $10^{-12}$, respectively.
All complex matrix entries were evolved directly.

% ------------------------------------------------------------
\subsection{Weak-defect prediction}
% ------------------------------------------------------------

At
\begin{equation}
k=1.5,
\end{equation}
the first-order formula gives
\begin{equation}
|A_T|^2
=
5.35601\times10^{-3},
\end{equation}
and
\begin{equation}
|A_R|^2
=
1.09610\times10^{-2}.
\end{equation}

Therefore
\begin{equation}
P_T^{1\rightarrow3}
\simeq
5.35601\times10^{-3}\delta_s^2,
\end{equation}
and
\begin{equation}
P_R^{1\rightarrow3}
\simeq
1.09610\times10^{-2}\delta_s^2
\end{equation}
for sufficiently small $\delta_s$.

% ------------------------------------------------------------
\subsection{Weak-defect scaling}
% ------------------------------------------------------------

Using eight odd-parity channels gives
\begin{table}[t]
\centering
\caption{Weak-defect scaling for the
$m=1$, $\ell=1\rightarrow3$ transition at $k=1.5$.}
\label{tab:weak_scaling}
\begin{tabular}{c|cc}
\hline
$\delta_s$
&
$P_T/\delta_s^2$
&
$P_R/\delta_s^2$
\\
\hline
0.0800
&
$5.11914\times10^{-3}$
&
$1.43226\times10^{-2}$
\\
0.0200
&
$5.32741\times10^{-3}$
&
$1.16743\times10^{-2}$
\\
0.0050
&
$5.35146\times10^{-3}$
&
$1.11346\times10^{-2}$
\\
0.00125
&
$5.35604\times10^{-3}$
&
$1.10064\times10^{-2}$
\\
\hline
\end{tabular}
\end{table}

As $\delta_s\rightarrow0$,
\begin{equation}
\frac{P_T}{\delta_s^2}
\longrightarrow
|A_T|^2,
\end{equation}
and
\begin{equation}
\frac{P_R}{\delta_s^2}
\longrightarrow
|A_R|^2.
\end{equation}

Thus the full coupled calculation approaches the predicted
\begin{equation}
P^{1\rightarrow3}
=
O(\delta_s^2)
\end{equation}
behavior.

% ------------------------------------------------------------
\subsection{Small-defect check}
% ------------------------------------------------------------

The weak-defect table gives a direct numerical check of the perturbative
result.  As $\delta_s$ is reduced, the ratios $P_T/\delta_s^2$ and
$P_R/\delta_s^2$ approach finite limits close to the coefficients obtained
from the first-variation formula.  We therefore use the scaling test, rather
than a separate finite-difference derivative, as the numerical verification
of Theorem~\ref{thm:scattering_variation}.

% ------------------------------------------------------------
\subsection{Angular-cutoff dependence}
% ------------------------------------------------------------

At
\begin{equation}
\delta_s=0.08,
\qquad
k=1.5,
\end{equation}
the dependence on the angular cutoff is
\begin{table}[t]
\centering
\caption{Dependence of the
$m=1$, $\ell=1\rightarrow3$ conversion probabilities on the angular
cutoff.}
\label{tab:angular_cutoff}
\begin{tabular}{c|cc}
\hline
$\ell_{\max}$
&
$P_T^{1\rightarrow3}$
&
$P_R^{1\rightarrow3}$
\\
\hline
3
&
$3.37612\times10^{-5}$
&
$9.50380\times10^{-5}$
\\
11
&
$3.28117\times10^{-5}$
&
$9.18269\times10^{-5}$
\\
15
&
$3.27625\times10^{-5}$
&
$9.16644\times10^{-5}$
\\
23
&
$3.27279\times10^{-5}$
&
$9.15509\times10^{-5}$
\\
\hline
\end{tabular}
\end{table}

Comparing $\ell_{\max}=15$ and $\ell_{\max}=23$, the relative changes
are approximately $1.1\times10^{-3}$ for transmission and
$1.2\times10^{-3}$ for reflection.

For $\ell_{\max}=23$, the total probability directly carried by
channels $\ell\geq5$ is
\begin{equation}
\sum_{\ell\geq5}
\left(
|T_{\ell1}|^2+|R_{\ell1}|^2
\right)
\simeq
7.07764\times10^{-8}.
\label{eq:higher_channel_probability}
\end{equation}

For the $\ell_{\max}=23$ calculation, the total outgoing flux differs
from unity by less than $10^{-10}$.

We also checked the size of the radial domain.  For $\ell_{\max}=15$,
changing the endpoint from $|l|=150$ to $|l|=200$ changes the transmitted
and reflected conversion probabilities by $8.3\times10^{-5}$ and
$7.5\times10^{-5}$ in relative terms.  Increasing the endpoint further
to $|l|=250$ changes them by only $2.6\times10^{-5}$ and
$1.4\times10^{-5}$, respectively.  The domain used for the tables is
therefore $|l|=200$.

These results support the stability of the low-channel calculation over
the truncations tested. They do not constitute a proof of convergence
of the full infinite-channel system.

% ------------------------------------------------------------
\subsection{No-ergoregion check}
% ------------------------------------------------------------

The stationary Killing field is
\begin{equation}
\xi=\partial_t.
\end{equation}
Its norm is
\begin{equation}
g_{tt}
=
-N^2
+
r^2\alpha^2\omega^2\sin^2\theta.
\end{equation}
Using
\begin{equation}
\omega=\frac{2J}{r^3},
\end{equation}
we obtain
\begin{equation}
g_{tt}
=
-N^2
+
\frac{4J^2\alpha^2}{r^4}
\sin^2\theta.
\end{equation}

The rotational term is largest at the equator. Therefore the
stationary Killing field remains timelike everywhere if
\begin{equation}
\boxed{
2|J|
<
\frac{N(l)r(l)^2}{\alpha(l)}
\qquad
\text{for every }l.
}
\label{eq:no_ergo_condition}
\end{equation}
Define
\begin{equation}
J_{\rm crit}
=
\frac{1}{2}
\min_{l\in\mathbb{R}}
\frac{N(l)r(l)^2}{\alpha(l)}.
\label{eq:Jcrit}
\end{equation}

For the numerical parameters used above, the minimum occurs at the
throat and
\begin{equation}
J_{\rm crit}
=
\frac{e^{-\Phi_0}r_0^2}
{2(1-\delta_s)}
=
0.40262\ldots .
\end{equation}
Since
\begin{equation}
J=0.05,
\end{equation}
the numerical calculation lies safely in the no-ergoregion regime.

No claim about gravitational or nonlinear stability is made.

% ============================================================
\section{Discussion}
\label{sec:discussion}
% ============================================================

The purpose of this paper is not to claim that a complete
cosmic-string-supported rotating wormhole has been constructed. The
geometry is prescribed, and the conical axes are singular. The aim is
more limited: to understand what follows when the conical factor in a
stationary axisymmetric wormhole is allowed to vary along the proper
radial direction.

The calculation gives two main consequences. One concerns the source
required by Einstein's equations. The other concerns scalar-wave
propagation. Both follow from the same function $\alpha(l)$.

% ------------------------------------------------------------
\subsection{A varying cone is not an ordinary variable-tension string}
% ------------------------------------------------------------

For a constant conical factor,
\begin{equation}
\alpha=\alpha_0,
\end{equation}
the local geometry near either polar axis has the familiar conical
form. It is tempting to replace the constant tension by
\begin{equation}
\mu(l)
=
\frac{1-\alpha(l)}{4G}.
\end{equation}

The mixed Einstein equation shows why this interpretation is
incomplete. When $\alpha'(l)\neq0$,
\begin{equation}
G_{l\theta}
=
-\frac{\alpha'}{\alpha}\cot\theta,
\end{equation}
and therefore
\begin{equation}
T_{\hat l\hat\rho}
=
-\frac{\alpha'}
{8\pi G\alpha\rho}
+
O(\rho).
\end{equation}
The corresponding integrated relation is
\begin{equation}
\lim_{\rho\rightarrow0}
2\pi\alpha\rho\,
T_{\hat l\hat\rho}
=
\mu'(l).
\end{equation}

Thus a longitudinally varying conical deficit requires an additional
exchange of momentum between the conical sector and some other source
sector. It is not simply an ordinary isolated ideal string with a
freely varying tension.

% ------------------------------------------------------------
\subsection{Two polar axes and the matter interpretation}
% ------------------------------------------------------------

Every fixed-$l$ angular section has two poles,
\begin{equation}
\theta=0,
\qquad
\theta=\pi.
\end{equation}
When $\alpha(l)\neq1$, both are conical.

Geometrically, it is therefore correct to refer to two polar conical
axes. It is stronger to interpret them as two independent physical
Nambu--Goto strings. Such an interpretation requires a matter model,
which is not constructed here.

For this reason, the source interpretation is made only after the
Einstein tensor has been calculated.

% ------------------------------------------------------------
\subsection{Scope of the near-axis obstruction}
% ------------------------------------------------------------

Theorem~\ref{thm:axis_obstruction} is an obstruction within the
restricted ansatz \eqref{eq:metric_general}. It does not prove that no
regular finite-width matter system can produce a conical deficit that
changes along a string-like core.

A regular core may modify the metric close to the axis, and additional
metric functions may become important there. The correct conclusion is
therefore more limited:
A regular finite-width realization therefore cannot be obtained by
keeping the restricted conical ansatz unchanged all the way to the
axis.

% ------------------------------------------------------------
\subsection{Different roles of the conical factor and its derivatives}
% ------------------------------------------------------------

The quantities
\begin{equation}
\alpha-1,
\qquad
\alpha',
\qquad
\alpha''
\end{equation}
play different roles.

The value of $\alpha-1$ determines the local conical deficit.

The derivative $\alpha'$ determines the mixed longitudinal--transverse
source through
\begin{equation}
T_{\hat l\hat\rho}
\sim
-\frac{\alpha'}{8\pi G\alpha\rho}.
\end{equation}

At a reflection-symmetric throat, $\alpha'(0)=0$, but $\alpha''(0)$
may be nonzero and contributes to the radial null contraction,
\begin{equation}
8\pi G T_{\mu\nu}k^\mu k^\nu
=
-2\frac{r''}{r}
-
\frac{\alpha''}{\alpha}.
\end{equation}

These effects are related but should not be identified with one
another.

% ------------------------------------------------------------
\subsection{Why scalar waves are part of the same problem}
% ------------------------------------------------------------

The same varying conical factor enters the Klein--Gordon equation
through
\begin{equation}
-\frac{m^2}
{r^2\alpha(l)^2\sin^2\theta}.
\end{equation}

For $m=0$, this term does not produce angular mixing. For $m\neq0$, a
nonconstant $\alpha(l)$ prevents ordinary radial-angular separation.

Thus the scalar problem asks what wave effect is produced by the same
longitudinal variation that gives the mixed Einstein constraint.

The answer is summarized by
\begin{equation}
\boxed{
\begin{array}{ccc}
\alpha'(l)\neq0
&\Longrightarrow&
\text{mixed Einstein source},
\\[2mm]
\alpha(l)\neq\text{constant}
&\Longrightarrow&
\text{angular mode coupling}.
\end{array}
}
\end{equation}

% ------------------------------------------------------------
\subsection{Meaning of the angular coupling}
% ------------------------------------------------------------

The coupling matrix
\begin{equation}
C^{(m)}_{j\ell}
=
\int_0^\pi
S_{jm}S_{\ell m}
\frac{d\theta}{\sin\theta}
\end{equation}
follows directly from the conical factor in the wave operator.

Its exact parity rule,
\begin{equation}
C^{(m)}_{j\ell}=0
\qquad
\text{for }j+\ell\text{ odd},
\end{equation}
shows that mode conversion is structured rather than arbitrary.

The bound
\begin{equation}
m^2C\leq\Lambda
\end{equation}
also shows that the singular angular coupling is controlled by the
ordinary angular energy, even though $C$ is not bounded on ordinary
channel $\ell^2$.

% ------------------------------------------------------------
\subsection{Meaning of the weak-defect scattering formula}
% ------------------------------------------------------------

For two different channels,
\begin{equation}
\frac{dS_{bj,a\ell}}
{d\delta_s}\bigg|_{\delta_s=0}
=
\frac{m^2C^{(m)}_{j\ell}}{ik}
\int_{-\infty}^{\infty}
\frac{N^2}{r^2}
f\,
u^{(0)}_{j,b}
u^{(0)}_{\ell,a}\,dx.
\end{equation}

The angular coefficient determines which transitions are allowed. The
radial integral determines how strongly the corresponding unperturbed
channels overlap in the region where the conical deformation is
present.

Since $f$ is localized near the throat, the conversion amplitude is
mainly sensitive to the wave functions in the throat region.

% ------------------------------------------------------------
\subsection{Role of rotation}
% ------------------------------------------------------------

Rotation is part of the prescribed background, but it is not the
source of the angular coupling studied here.

The frame-dragging function enters the scalar radial equation through
\begin{equation}
(\sigma-m\omega)^2,
\end{equation}
whereas the off-diagonal channel coupling comes from
\begin{equation}
m^2(\alpha^{-2}-1)C^{(m)}.
\end{equation}

Likewise,
\begin{equation}
G_{l\theta}
=
-\frac{\alpha'}{\alpha}\cot\theta
\end{equation}
is independent of $\omega$.

Therefore the varying-cone source obstruction and the angular coupling
are already present in the static case. Rotation changes the radial
wave propagation and the causal structure, but it is not responsible
for these two effects.

% ------------------------------------------------------------
\subsection{Limitations}
% ------------------------------------------------------------

The model has several clear limitations.

First, the spacetime is prescribed rather than obtained by solving
Einstein's equations together with a regular finite-width matter
action.

Second, the polar axes remain singular when the metric is extended all
the way to $\rho=0$ with nonconstant $\alpha$.

Third, the scattering formula is derived in the weak-defect regime.

Fourth, the numerical calculation uses finite angular truncations. The
form estimate controls the angular perturbation, but a full proof of
convergence of the infinite-dimensional scattering matrix has not been
given.

Finally, only a minimally coupled test scalar is considered. No claim
about gravitational or nonlinear stability is made.

% ============================================================
\section{Conclusion}
\label{sec:conclusion}
% ============================================================

We have studied a stationary axisymmetric wormhole whose conical
factor varies with the proper radial coordinate,
\begin{equation}
\alpha=\alpha(l).
\end{equation}

The first result concerns the source required by the geometry. For the
metric class considered here,
\begin{equation}
G_{l\theta}
=
-\frac{\alpha'}{\alpha}\cot\theta.
\end{equation}
Near either polar axis,
\begin{equation}
T_{\hat l\hat\rho}
=
-\frac{\alpha'}
{8\pi G\alpha\rho}
+
O(\rho).
\end{equation}
Therefore a nonconstant conical factor cannot be supported by matter
that remains bounded near the axis while the restricted metric ansatz
is kept unchanged.

The integrated relation
\begin{equation}
\lim_{\rho\rightarrow0}
2\pi\alpha\rho\,
T_{\hat l\hat\rho}
=
\mu'(l),
\qquad
\mu(l)=\frac{1-\alpha(l)}{4G},
\end{equation}
shows that a varying deficit requires additional source structure. It
should not be interpreted simply as an ordinary ideal cosmic string
with a freely varying tension.

The second result concerns scalar waves. For $m=0$, the wave equation
does not produce angular-channel mixing. For $m\neq0$, a nonconstant
$\alpha(l)$ destroys ordinary product separation. The angular coupling
matrix can be evaluated in closed form and satisfies
\begin{equation}
C^{(m)}_{j\ell}=0
\qquad
\text{for }j+\ell\text{ odd}.
\end{equation}
It also obeys the quadratic-form estimate
\begin{equation}
m^2C\leq\Lambda.
\end{equation}

In the weak-defect regime, the first variation of an off-diagonal
scattering coefficient is
\begin{equation}
\frac{dS_{bj,a\ell}}
{d\delta_s}\bigg|_{\delta_s=0}
=
\frac{m^2C^{(m)}_{j\ell}}{ik}
\int_{-\infty}^{\infty}
\frac{N^2}{r^2}
f\,
u^{(0)}_{j,b}
u^{(0)}_{\ell,a}\,dx.
\end{equation}
Thus the conversion amplitude is first order in $\delta_s$, while the
probability begins at order $\delta_s^2$.

For the first allowed transition,
\begin{equation}
m=1,
\qquad
\ell=1\longrightarrow3,
\end{equation}
the numerical coupled-channel calculation approaches the analytic
weak-defect prediction as $\delta_s\rightarrow0$, and the low-channel
result is stable over the angular truncations tested.

The paper does not construct a regular finite-width matter source and
does not prove convergence of the complete infinite-channel scattering
problem. These are separate questions.

The main conclusion is therefore limited but clear: a conical deficit
that varies along the wormhole changes both the required source and
the scalar-wave dynamics.

% ============================================================
\section*{Data availability}
% ============================================================

All numerical data supporting the results reported in this article are
contained in the tables and equations above.  The numerical implementation
used to reproduce the finite-channel calculations is supplied with the
submission as supplementary material.

% ============================================================
% References
% ============================================================

\end{document}